\input epsf
\magnification=\magstep1 
\overfullrule=0pt
\parskip=3pt 
 at 14truept
\overfullrule 0pt


\catcode`\@=11
\font\tenmsa=msam10
\font\sevenmsa=msam7
\font\fivemsa=msam5
\font\tenmsb=msbm10
\font\sevenmsb=msbm7
\font\fivemsb=msbm5
\newfam\msafam
\newfam\msbfam
\textfont\msafam=\tenmsa  \scriptfont\msafam=\sevenmsa
  \scriptscriptfont\msafam=\fivemsa
\textfont\msbfam=\tenmsb  \scriptfont\msbfam=\sevenmsb
  \scriptscriptfont\msbfam=\fivemsb

\def\hexnumber@#1{\ifcase#1 0\or1\or2\or3\or4\or5\or6\or7\or8\or9\or
        A\or B\or C\or D\or E\or F\fi }

\def\Bbb{\ifmmode\let\next\Bbb@\else
 \def\next{\errmessage{Use \string\Bbb\space only in math mode}}\fi\next}
\def\Bbb@#1{{\Bbb@@{#1}}}
\def\Bbb@@#1{\fam\msbfam#1}

\def\\{\hfil\break}
\def\la{\lambda}

\def\om{\omega}
\def\si{\sigma}

\def\l{\ell}

\def\D{{\cal D}}
\def\Q{{\cal Q}}
   
\def\al{\alpha}

\def\Q{{\cal Q}} 

\def\b#1{\big(#1\big)}

\def\+{\oplus}

\font\huge=cmr10 scaled \magstep2
\font\small=cmr8


{\nopagenumbers
\rightline{September, 1995}
\rightline{hep-th/9511089}
\vskip2cm
\centerline{{\huge\bf Spectra of Conformal Field Theories with 
Current Algebras\footnote{$^\ddagger$}{\small Based on a talk
given by MW at the CRM-CAP Workshop in Theoretical and Mathematical
Physics, U. Laval, Qu\'ebec, 6/95}}} \bigskip\bigskip\centerline{T.
Gannon$^\bullet$\footnote{$^*$}{\small  E-mail:
gannon@mpim-bonn.mpg.de},   P. Ruelle$^\circ$\footnote{$^\star$}{\small
Chercheur Qualifi\'e FNRS. E-mail: ruelle@fyma.ucl.ac.be} and M.A.
Walton$^\odot$ {\footnote{$^\dagger$}{\small Supported in part by NSERC.
E-mail: walton@hg.uleth.ca}}} \bigskip
\centerline{$^\bullet${\it Max-Planck-Institut f\"ur Mathematik}} 
\centerline{{\it Bonn, Germany, 53225}}
\bigskip
\centerline{$^\circ${\it Institut de Physique Th\'eorique et Math\'ematique,}} 
\centerline{\it Universit\'e Catholique de Louvain, Louvain--la--Neuve, Belgium} 
\bigskip
\centerline{$^\odot${\it Physics Department,
University of Lethbridge}} \centerline{{\it Lethbridge, Alberta, 
Canada\ \ T1K 3M4}} 

\vskip1cm \leftskip=2cm \rightskip=2cm
\noindent{{\bf Abstract:}} 
This is an elementary review of our recent work on the classification
of the spectra  of those two-dimensional rational conformal field theories
(RCFTs) whose (maximal) chiral algebras are  
current algebras. We classified all possible partition functions for such
theories when the defining finite-dimensional Lie algebra is simple. The
concepts underlying this work are emphasized, and are illustrated using simple 
examples.  

\leftskip=0cm \rightskip=0cm

\vfill 

\eject}

\pageno=1

Perhaps the most fundamental question one can ask about conformal field
theories is ``which exist?''. More precisely, fixing a chiral algebra, we
would like to know the possible spectra of primary fields in any consistent
theory. The possible spectra determine the universality classes of critical
two-dimensional systems, and also restrict any critical string
theory realizing the particular chiral algebra. 

As Cardy [1] pointed out, spectra are very strongly constrained
by the modular invariance of the torus partition function. For
example, when the chiral algebra contains the su(2) current algebra (or
nontwisted affine Kac-Moody algebra) and there is a finite number of current
algebra primary fields, Cappelli, Itzykson and Zuber [2] were able to classify
the possible spectra using modular invariance. The analogous exercise for the
next-simplest simple Lie algebra, su(3), was only completed much later, by one
of us [3]. Recently, however, we were able to give the complete list of
possible spectra for the special case when the {\it maximal} chiral algebra is
a current algebra based on {\it any} simple Lie algebra [4]. This work
followed [5], where the classification was completed for the simplest series of
simple Lie algebras, the $A_\l$ (see [4] for further references on
modular invariants and classification of spectra).

Here we present an elementary review of [4], emphasizing the concepts
involved with simple illustrative examples. Let us first state the problem
precisely. 

Let $X_\l$ be a finite-dimensional simple Lie algebra, and let $X_{\l,k}$
denote the corresponding current algebra, at positive integer level $k.$ The
torus partition function of a conformal field theory containing a finite number
of $X_{\l,k}$ primary fields may be expressed as a sesquilinear combination of
its characters: 
$$
Z = \sum_{\mu,\mu' \in {\bar P}_+(X_{\l,k})} \, 
M_{\mu,\mu'} \chi^*_\mu \, \chi_{\mu'}\ \ .
\eqno(1)
$$
$M$ is a non-negative integer matrix with $M_{0,0}=1$, and $\chi_\lambda$
denotes the character of the integrable representation of $X_{\l,k}$ of highest
weight corresponding to dominant $X_\l$ weight $\la.$ The set of integrable
highest weights of $X_{\l,k}$ is in one-to-one correspondence with the
following set of dominant weights of $X_\l$:$$ {\bar P}_+(X_{\l,k}) = \big\{
\la=(\la_1,\la_2,\ldots,\la_\l) \;|\; \la_i \in {\Bbb N} \hbox{ and }
\sum_{i=1}^\ell a_i^\vee \la_i \le k \big\}\ , \eqno(2)
$$
where the $a_i^\vee$ are the colabels of $X_\l$. For short, we will refer to
this set as the alc\^ove. 

If we restrict to the case when the current algebra is the {\it maximal} chiral
algebra, then $M$  defines
a permutation $\si$ of the alc\^ove by $M_{\mu,\mu'} = \delta_{\mu',\si
(\mu)}$ [6]. The characters $\chi_\la(\tau,z,u)$ transform linearly under the
action of the modular group, with generators 
$T:\ (\tau,z,u) \mapsto (\tau+1,z,u)$ and $S:\ (\tau,z,u) \mapsto
({-1 \over \tau},{z \over \tau},u+{z^2\over 2\tau})$. The representing
matrices take the form [7] 
$$\eqalignno{
& T_{\la,\la'} = \gamma \exp{\left( {2\pi i (\rho+\la)^2 \over 2n} \right)} \,
\delta_{\la,\la'}, &(3a) \cr
& S_{\la,\la'} = \gamma' \sum_{w \in W} \,({\rm det}\,w) \exp{ \left(
-{2\pi i (\rho +\la) \cdot w(\rho + \la') \over n} \right)}, &(3b) \cr}
$$
where $n=k+h^\vee,$ $h^\vee$ being the dual Coxeter number of $X_\l,$ and
$\rho=\sum_{i=1}^\l \om^i$ is the Weyl vector, the sum of fundamental weights
$\om^i$. $\gamma$ and $\gamma'$ are constants independent of $\la$ and $\la'$,
and $W$ is the Weyl group of $X_\l$. The matrices $S$ and $T$ are both
symmetric and unitary, and satisfy $S^2=(ST)^3=C$. $C$ is charge conjugation, 
an order 2 symmetry of the Coxeter-Dynkin diagram of $X_\l$ (if non--trivial).

Modular invariance of (1) demands that
$\si$ commutes with the matrices $S$ and $T$, i.e.
$$\eqalignno{
& T_{\la,\la'} = T_{\si(\la),\si(\la')}, &(4a) \cr
& S_{\la,\la'} = S_{\si(\la),\si(\la')}. &(4b) \cr}
$$
Any permutation of $X_{\l,k}$ obeying (4a,b) is called an automorphism
invariant. In [4], we classified all such automorphism invariants, for any
simple Lie algebra $X_\l$. 

Condition (4a), $T$-invariance, is easy to apply, since $T$ is the 
simple, diagonal matrix of (3a). Not simple to apply directly is the condition
(4b) sufficient for $S$-invariance, but one can find necessary (but not
sufficient) conditions that are easy to impose yet still are strong
restrictions. First, consider Verlinde's formula for the
fusion coefficients:
$$
N_{\la,\mu}^\nu = \sum_{\beta \in {\bar P}_+(X_{\l,k})} \,
{S_{\la,\beta} S_{\mu,\beta} S^*_{\nu,\beta} \over S_{0,\beta}}.
\eqno(5)
$$
Now, because $S_{\la,\mu}>0$ for all $\mu$ only if $\la=0,$ we can show that
$\si(0)=0$ follows from (4b). 
Applying this and (4b) to Verlinde's formula gives
$$
N_{\si(\la),\si(\mu)}^{\si(\nu)} = N_{\la,\mu}^\nu\ \ ,
\eqno(6)
$$
i.e. any automorphism modular invariant is a symmetry of the fusion
coefficients. Since the fusion coefficients $N_{\la,\mu}^\nu$ are non-negative
integers and are more easily calculable than the matrix $S$ (see [8] and
references therein), this last condition proved easier to impose than (4b). 

The second condition necessary for (4b) that we used involved the so-called
quantum dimensions, defined by
$$
\D(\la) :={S_{0,\la} \over S_{0,0}} = \prod_{\alpha >0} \;
{\sin{[\pi \alpha \cdot (\rho +\la)/n]} \over 
\sin{[\pi \alpha \cdot \rho /n]}}\ .
\eqno(7)
$$
It is easy to see that
$$
\D(\si\la)\ =\ \D(\la)\ \ .
\eqno(8)
$$
Because the product formula (7) for $\D$ is considerably simpler than the
Weyl-sum formula (3b) for $S$, the condition (8) is more manageable than
the full $S$-invariance (4b).

\epsfxsize=13cm
\epsfbox{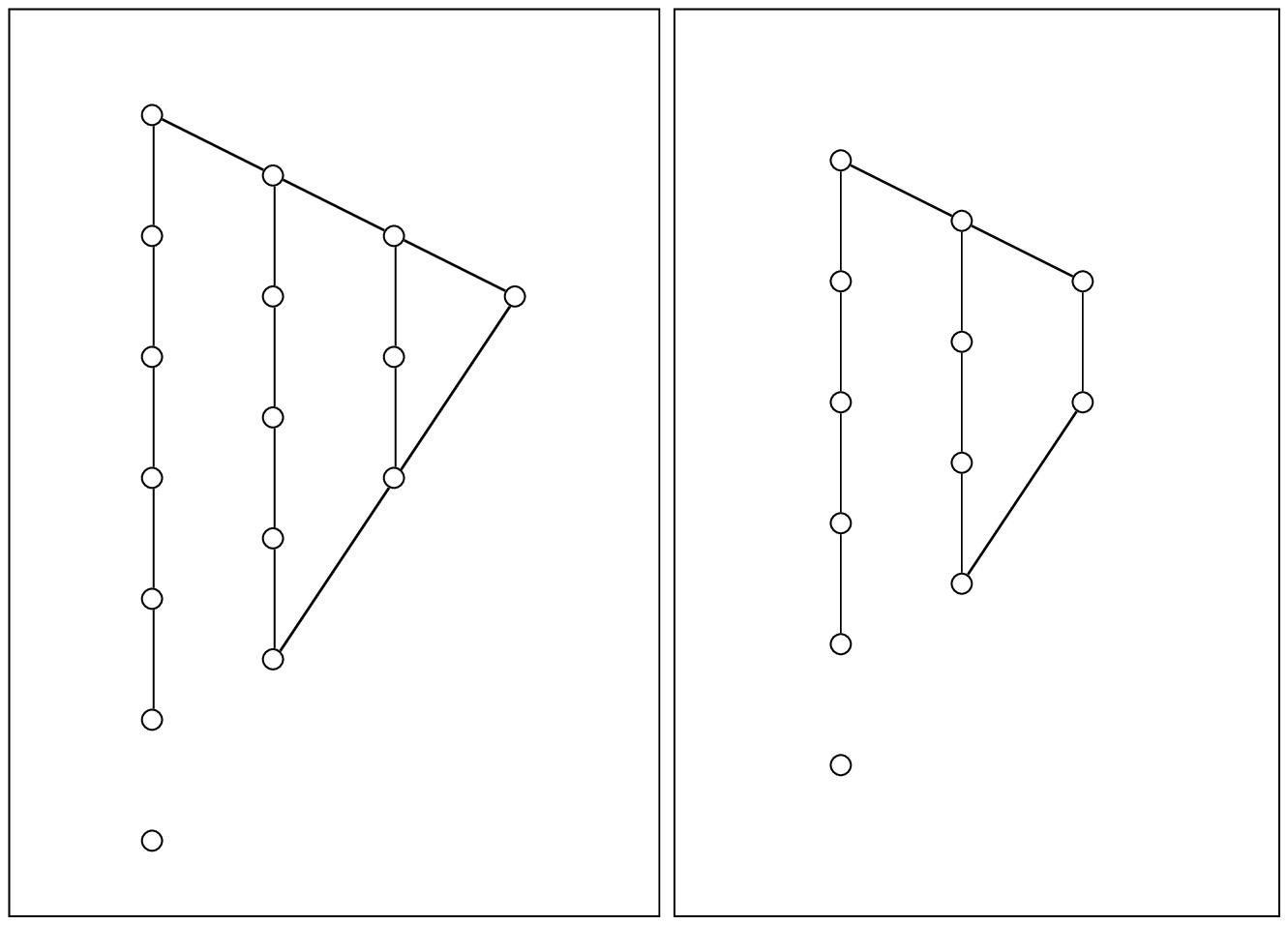}
\noindent{{\it Figures 1 and 2}. Circles indicate the weights of the alc\^oves
of $G_{2,6}$ and $G_{2,5}$, in Figures 1 and 2, respectively. As explained in
the text, the lines drawn rule out all but a few candidates for weights of
second-lowest quantum dimension.} \vskip.5cm 

Now, any automorphism invariant is determined by its action on a subset
of the weights of the alc\^ove. In
particular, one can construct an automorphism invariant from its action on the 
fundamental weights in the alc\^ove [4]. Our strategy was to find all possible
automorphism invariants by finding all possible actions on the fundamental
weights that are consistent with (4a), (6) and (8).

First, the sets ${\cal Q}_2$ of weights having second-smallest quantum
dimension were found. (The sets ${\cal Q}_1$ were previously determined by
Fuchs [9], and we made use of some of his ideas.) By (8), an automorphism
invariant can only permute a weight in ${\cal Q}_2$ to itself,
or another weight in ${\cal Q}_2.$ By checking T-invariance (4a), the list of
possibilities could be immediately reduced. ${\cal Q}_2$ always included a
fundamental weight (or some other simple weight), $\omega^f$ say, so that only
a small number of possibilities for $\sigma\omega^f$ could be listed. By
studying fusion rules involving $\omega^f$ and the other fundamental weights,
(6) could then be used with (4a) to list  all possible actions on fundamental
weights, and therefore the possible full automorphism invariants. 

It was important that for fixed $X_{\l,k}$, the set of automorphism invariants
forms a {\it group}, with composition as the group multiplication. This meant
that known automorphism invariants could be ``factored off'', i.e. it was often
possible to simplify a potential
automorphism invariant $\tilde\si$ by
multiplying by a known automorphism invariant $\si$. For example, suppose we
know that $\tilde\si\la=\mu,$ and a known automorphism invariant $\si$
also obeys $\si\la=\mu.$ Then, if $\tilde\si$ is an automorphism
invariant, so will be $\si^\prime:=\si^{-1}\circ\tilde\si,$ but with
$\si^\prime\la=\la.$ This proved very useful, since many automorphism invariants
were already known (see the references of
[4]). 
  
Henceforth we will concentrate on the simple rank-two examples of $X_\l=G_2$
and $C_2,$ to illustrate our proof. The first task is to find $\Q_2$ for
these algebras. If we consider the quantum dimension as a function of weights
with real Dynkin labels $\la_i$, the function is quite simple, as it turns
out. Along any straight line in weight space that begins and
ends on weights of the alc\^ove, the minimum value can only occur at the ends
[4]. This reduces the possible elements of $\Q_2$ enormously.

For example, consider the alc\^ove of $G_{2,6}$ drawn in Figure 1. For $G_2$,
$\Q_1=\{0\},$ for all levels $k.$ By considering the lines indicated, one can
show that $\om^1,\ \om^2,\ 6\om^2,\ 3\om^1$ are the only possible candidates
for elements of $\Q_2.$ For odd levels, the situation is slightly different,
as Figure 2 shows. There the case $G_{2,5}$ is drawn, and the candidates are 
$\om^1,\ \om^2,\ 5\om^2,\ 2\om^1,\ 2\om^1+\om^2.$ Clearly then, the candidates for even
$k$ are just $\om^1,\ \om^2,\ k\om^2,\ {k\over 2}\om^1$, and those for odd
level $k$ are $\om^1,\ \om^2,\ k\om^2,\ {k-1\over 2}\om^1, {k-1\over
2}\om^1+\om^2.$

Figure 3 similarly indicates the example of $C_{2,3}$. There is the
complication of a diagram automorphism, and corresponding simple
current [10] $J$ in the case of $C_2.$ The simple current $J$ acts as a
reflection on the weights of the alc\^ove, as indicated. Since
$\D(\la)=\D(J\la),$ we can only specify the $J$-orbits $[\la]$ of the
candidates $\la$.  Using $\Q_1=[0]=\{0,k\om^2\},$ and by considering the lines
drawn in Figure 3, it is easy to see that the candidates for $C_{2,k}$ are
contained in $[\om^1],\ [\om^2]$ and $[k\om^1].$

The candidates $\la$ for elements of $\Q_2$ come in two varieties: those that
are independent of the level $k,$ and those that have a single Dynkin label
that grows linearly with $k.$ The $k$-independent ones have quantum
dimensions that tend to the ordinary (Weyl) dimension of the corresponding
$X_\l$ representation in the limit of large $k.$  For large enough $k$ then,
we expect the weight $\la$ that is the highest weight of the
second-lowest-dimensional representation to be the only surviving candidate
of this type. In fact, from (7) follows 
$$
{\partial\over \partial k}\log\left({\D(\la) \over \D(\mu)}\right) = 
{\pi\over n^2}\sum_{\al>0}\Big[(\mu+\rho)\cdot\al\,\cot\Big(
\pi\,{(\mu+\rho)\cdot\al\over n}\Big)-(\la+\rho)\cdot\al\,\cot\Big(\pi\,{
(\la+\rho)\cdot\al\over n}\Big)\Big]\ .
\eqno(9)
$$
For $G_2$ with $\la=\om^1$ and $\mu=\om^2,$ the right hand side of (9) is
positive for all levels $k\ge 1.$ 
Numerically, we find that $\D(\om^1)=\D(\om^2)$ when $k=3,$ so (9) tells
us that $\D(\om^1)>\D(\om^2)$ for all levels $k\ge 4.$ Similarly, for $C_2$
with  $\la=\om^2$ and $\mu=\om^1,$ the right hand side is positive for all
levels. Again, numerically we find that $\D(\om^1)=\D(\om^2)$ when $k=3.$ For
$C_2$ then, $\D(\om^2)>\D(\om^1)$ for all levels $k\ge 4.$ For low levels ($\le
3$) the set $\Q_2$ can be found easily numerically, but for higher levels ($k\ge
4$), we still must compare the quantum dimensions of $\om^2$ for $G_2$ and
$\om^1$ for $C_2$ to those of candidates of the second variety.
\vfill\eject

\vskip-1cm
\epsfxsize=13cm
\epsfbox{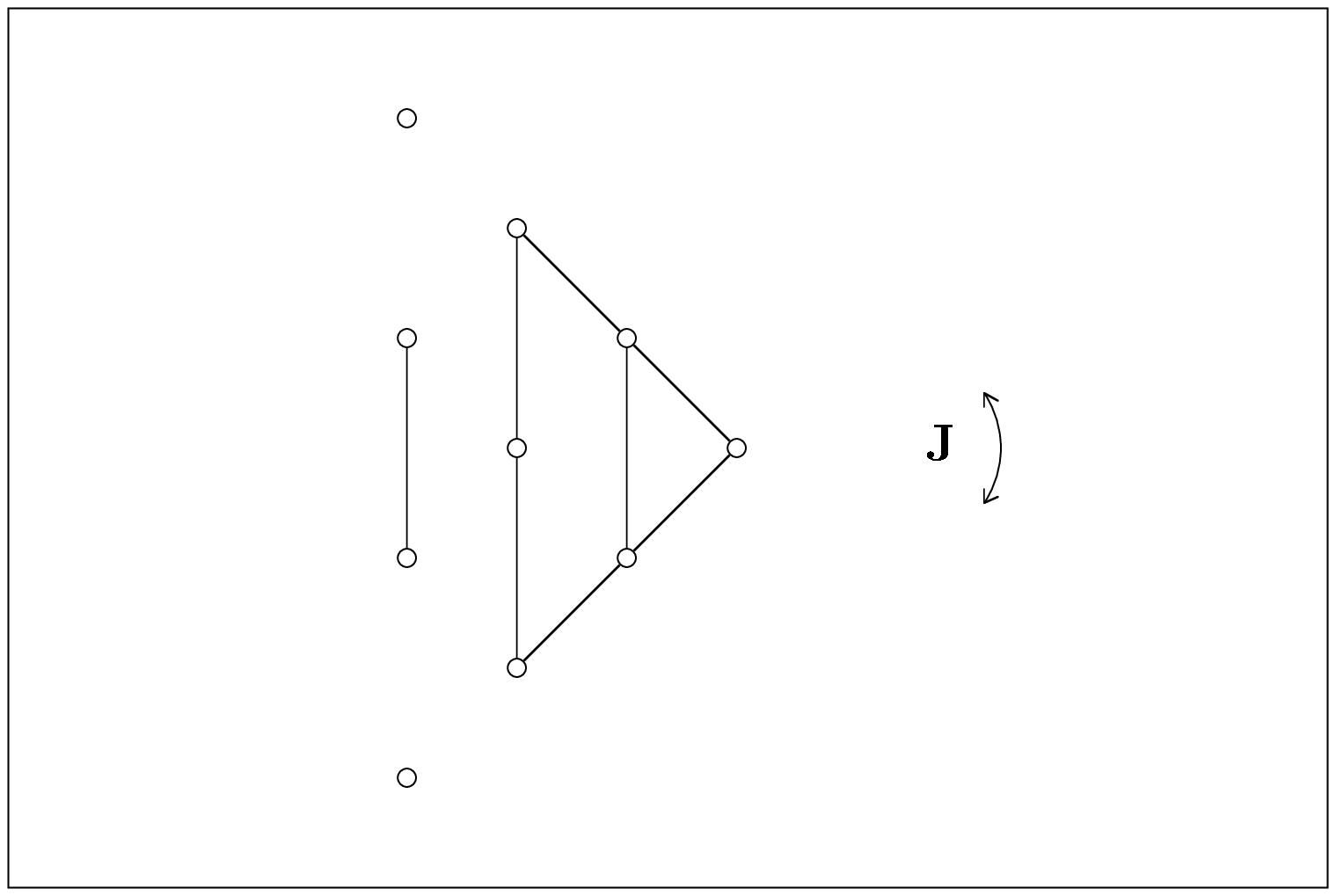}
\noindent{{\it Figure 3}. Candidates for $C_{2,3}$ weights of second-lowest
quantum dimension. The simple current $J$ reflects the weights as shown.}
\vskip.5cm

Consider $C_2$ first. We will use the well-known level-rank duality of quantum
dimensions. Suppose $\la$ is a weight in the alc\^ove of $C_{\l,k}$, and
$\la^\prime$ is the weight in the alc\^ove of $C_{k,\l}$ whose Young tableau is
the transpose of that of $\la.$ Then $\D(\la)=\D^\prime(\la^\prime),$ where
$\D$ indicates a $C_{\l,k}$ quantum dimension, and $\D^\prime$ one for
$C_{k,\l}.$ For the algebra $C_{k,2}$ dual to $C_{2,k}$, one can show that 
$\D'(\om'^k) > \D'(\om'^1)$ for all $k\ge 4,$ in a manner similar to the
argument of the previous paragraph. Therefore, rank-level duality
tells us that
$$
\D(k\om^1) = \D'(\om'^k) > \D'(\om'^1) = \D(\om^1).
\eqno(10)
$$
So, for all levels $k\ge 4,$ we find $\Q_2=[\om^1].$ The lower levels can be
treated numerically, with results: k=1, $\Q_2=[\om^1]$; $k=2$,
$\Q_2=[\om^2]\cup [2\om^1]$; $k=3$, $\Q_2=[\om^1]\cup [\om^2]\cup [3\om^1]$.

The situation is more complicated for $G_2$; for example, there is no
level-rank duality available. However, the quantum dimension of a k-dependent
candidate $\la^k$ is a monotonically increasing function of $k$ for large enough
level. Now, the quantum dimension of the surviving $k$-independent candidate
$\om^f$  ($\om^2$ for $G_2$) is also an increasing function of $k$, but one that
converges to its Weyl dimension $dim(\om^f)$. Therefore, if for a certain level 
$k=k_1$ we have $\D(\la^{k_1})\ge dim(\om^f)$, then $\D(\la^k)>\D(\om^f)$ is 
guaranteed for all higher levels $k>k_1.$ 

It is not difficult to show directly from (7) that all k-dependent candidates
for $G_2$ have quantum dimensions that are increasing functions of $k$ for $k\ge
2$. Numerically, we found that they all also exceeded
$dim(\om^f)=dim(\om^2)=7$ for $k\ge 6$. Therefore, the $k$-dependent
candidates can be eliminated, so that $\Q_2=\{\om^2\}$ for all levels $k\ge
6.$ Treating the low levels numerically, we finally find $\Q_2=\{\om^2\}$ for
$k=1$ and all $k\ge 5,$ $\Q_2=\{\om^1\}$ for $k=2$,
$\Q_2=\{\om^1,\om^2,3\om^2\}$ for $k=3$, and $\Q_2=\{\om^2,2\om^1\}$
for $k=4$. 

The weights of second-lowest quantum dimension for both $C_2$ and $G_2$ have
now been listed. Now we must select a fundamental field, or some other
simple field, $\om^f\in \Q_2$, and restrict the possibilities $\si\om^f\in
\Q_2$ using $T$-invariance, eqn. (4a). To convey the method, it will be
sufficient to restrict to particular levels, and consider just $G_{2,4}$ and
$C_{2,3}$. 

Level $k=4$ is the most interesting for $G_2$. Choosing $\om^f=\om^2,$ we see
that  $\si\om^f=\om^f$ and $\si\om^f=2\om^1$ are the only possibilities, since 
$\Q_2=\{\om^2,2\om^1\}$. From the Kac-Peterson formula for the modular matrix
$T$, eqn. (3a), $T$-invariance means that if $\si\la=\mu,$ we must have 
$(\la+\rho)^2\equiv(\mu+\rho)^2({\rm mod}\ 2(k+h^\vee))$. With $\la=\om^2$ and
$\mu=2\om^1$ this yields $8{2\over 3}\equiv 24{2\over 3}({\rm mod}\ (2k+8))$.
This means $\si\om^f=2\om^1$ is possible {\it only} for level $k=4.$ 

So, for $G_{2,4}$ there remain two possibilities: either $\si\om^2=\om^2,$ or
$\si\om^2=2\om^1$. The first possibility implies that $\si=id$. To see this,
we use the symmetry of the fusion coefficients (6) on the following fusion
rule:
$$
\om^2 \otimes \om^2 = \b{0}^{4{2 \over 3}}_1 \;\+\; \b{\om^1}^{12{2 \over 3}}_2
\;\+\; \b{\om^2}^{8{2 \over 3}}_1 \;\+\; \b{2\om^2}^{14}_2\,.
\eqno(11)
$$
We write the fusion rule for all levels as a single tensor product
decomposition for $G_2,$ with subscripts indicating the {\it threshold level} 
of the fusions. For example, (11) indicates that the representation of highest
weight $2\om^2$ is contained in the fusion product $\om^2 \times \om^2$ for
all levels greater than or equal to 2. For convenience, the values
$(\la+\rho)^2$ for highest weight $\la$ are also indicated, as superscripts.
By (6), if $\si\om^2=\om^2$, the right hand side of (11) must be invariant
under the action of $\si.$ The only nontrivial possibility is that $\si$
interchanges $\b{\om^1}^{12{2 \over 3}}_2$ and $\b{2\om^2}^{14}_2.$ But
$T$-invariance demands that the superscripts be equivalent (mod
$2(k+h^\vee)=16$). Therefore, $\si\om^1=\om^1$ as well as $\si\om^2=\om^2.$
Since $\si$ fixes all the fundamental weights of $G_{2,4}$, we conclude
$\si=id.$

The other possibility, $\si\om^2=2\om^1$ is realized in an automorphism
invariant of the Galois type first found by Verstegen. We denote
this invariant $\si_{g2}$. As illustrated in Figure 4, it interchanges the
weights $\om^2$ and $2\om^1$, and the pair $\om^1,\ 4\om^2$, and fixes all
others. It is simple to show that $\si_{g2}$ is the {\it only} invariant
sending $\om^2$ to $2\om^1$, using the group property of automorphism
invariants. For $\si_{g2}^{-1}\circ\si$ fixes $\om^2,$ and we showed in the
previous paragraph that any such invariant must be the identity.

Therefore, the full set of invariants for $G_{2,4}$ is just
$\{id,\si_{g2}\}$. For all other levels $k\ge 1,$ we found that the identity
is the unique automorphism invariant for $G_2$.

For $C_{2,3}$ we found above that $\Q_2=[\om^1]\cup [\om^2]\cup [3\om^1]$.
More explicitly,
$\Q_2=\{(\om^1)^5,(\om^1+2\om^2)^{17},(\om^2)^{13/2},(2\om^2)^{25/2},(3\om^1)^{13}\},$
where the superscripts again indicate the ``norm squared'' $(\la+\rho)^2$. 
Since $2(k+h^\vee)=12$ for $C_{2,3}$, $T$-invariance tells us that the only
possibilities are $\si\om^1=\om^1,$ or $\si\om^1=\om^1+2\om^2$. The former
possibility is consistent only with $\si=id,$ as can be seen from the
following fusion rule:
$$
\om^1 \times \om^1 = (0)^{5/2} + (\om^2)^{13/2} + (2\om^1)^{17/2}\ \ .
\eqno(12)
$$
If $\si\om^1=\om^1,$ then the right hand side of (12) is fixed by $\si$. But
comparison of the superscripts shows that only $\si\om^2=\om^2$ is consistent
with $T$-invariance. Therefore both fundamental weights are fixed, so that the
automorphism invariant must be the identity.

\vskip-1cm
\epsfxsize=16cm
\epsfbox{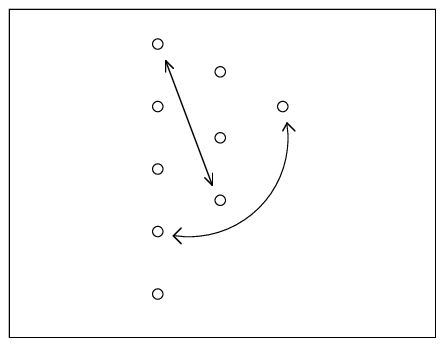}
\vskip0cm
\noindent{{\it Figure 4.} Indicated is $\si_{g2}$, the $G_{2,4}$ automorphism
modular invariant of Galois type. Its action on the weights of the
alc\^ove is shown using arrows. The weights $\om^2$ and $2\om^1$ are
interchanged, as are $\om^1$ and $4\om^2$; all others are fixed.}\vskip.5cm 

The other possibility, $\si\om^1=\om^1+2\om^2$, is realized in an invariant of
the simple current type, illustrated in Figure 5. But if we denote this as
$\si_{J},$ then $\si_{J}^{-1}\circ\si$ fixes $\om^1$. Therefore it
also fixes $\om^2$ by the argument above, and so must be the identity. We
conclude that the only invariants for $C_{2,3}$ are the identity, and one
simple current invariant, pictured in Fig. 5.

The classification of all automorphism modular invariants of current algebras
$X_{\l,k}$ for all simple $X_\l$ proceeds essentially in the same manner, but
with more complications. To conclude, we'll comment on our
findings [4]. 

No new surprises were
found; all modular invariants listed were previously known. The existence of
infinite series of modular invariants of the orthogonal algebras ($B_\l$ and
$C_\l$) at level 2 was only recently uncovered [11], however.  In [4] we gave
the first explicit, direct formulas for these invariants. Our formulas make it
clear that they are of a new type (dubbed generalized Galois
invariants) that also include invariants in the more general context of
rational conformal field theory. 

Our catalogue of automorphism invariants contains invariants constructible by
the orbifold procedure, the so-called simple current invariants, and those
constructible using the Galois properties of modular matrices [11]. Along
with the new generalized Galois invariants, there is also the very
exceptional $E_{8,4}$ invariant first discovered by Fuchs and Verstegen. It is
neither a simple current invariant nor a Galois (nor generalized Galois)
invariant. A natural question then is ``does there exist a method of
constructing modular invariants that includes the $E_{8,4}$ in a more general
class?''.

\vskip9cm
\epsfxsize=7cm
\epsfbox{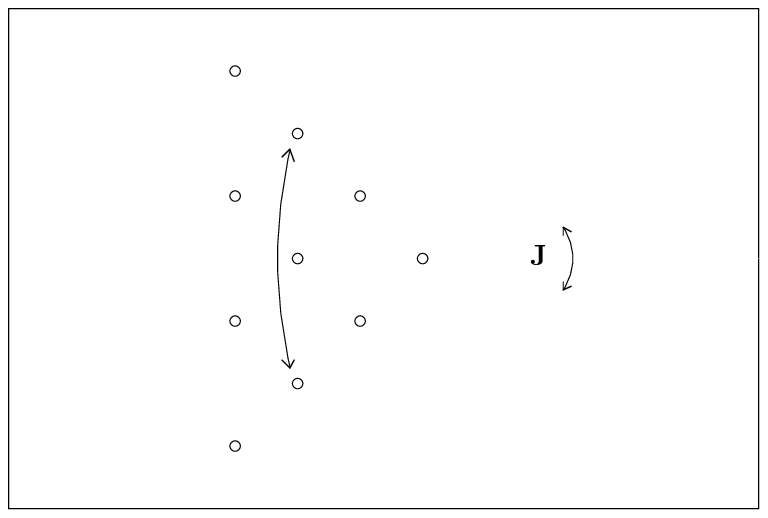}
\vskip-4cm
\noindent{{\it Figure 5}. Shown is the action of the simple current invariant
$\si_J$ on the alc\^ove of $C_{2,3}$.}\vfill\eject

Of course, one such method of constructing invariants is that implied by
our classification proof. But this procedure is not very
direct, although it {\it can} in principle be applied to rational conformal
field theories  other than those realizing current algebras.

Modular invariants corresponding to conformal field theories realizing 
{\it extensions} of chiral algebras, the extension invariants, are not
amenable to our method. For example, the group property of automorphism
invariants does not carry over to extension invariants. But perhaps some
variation will work. In a sense, however, it is the automorphism invariants
that are the most natural. As soon as the chiral algebra is extended, a
better description of the theory would be as an automorphism invariant of the
extended algebra [6]. On the other hand, the famous A-D-E pattern of
$A_{1,k}$ invariants [2] only appears when both automorphism and extension
invariants are included. No pattern of a similar nature is (as yet)
transparent in our classification. 
    
\vskip.5cm
\noindent{\it Acknowledgements.}  TG acknowledges the IHES and the Concordia
University Math Dept for their hospitality. MW thanks the organizers of the
CRM-CAP Workshop for the invitation that allowed him to present this material. 

\vskip.5cm \noindent{\bf References.}

\vskip0.25cm 

\item{1.} Cardy, J.: The operator content
of two-dimensional conformally invariant theories. Nucl. Phys. {\bf B270},
186-204 (1986)

\item{2.} Cappelli, A., Itzykson, C., Zuber, J.-B.: The A-D-E
classification of $A_1^{(1)}$ and minimal conformal field theories. Commun.
Math. Phys. {\bf 113}, 1-26 (1987)

\item{3.} Gannon, T.: The classification of affine $su(3)$ modular
invariant partition functions. Commun. Math. Phys. {\bf 161}, 233-264 (1994)

\item{4.} Gannon, T., Ruelle, P., Walton, M.A.: Automorphism invariants of
current algebras. Commun. Math. Phys., to appear (1995)

\item{5.} Gannon, T.: Symmetries of the Kac-Peterson modular matrices
of affine algebras. To appear in Inv.\ Math.

\item{6.} Moore, G., Seiberg, N.: Naturality in conformal field theory.
Nucl. Phys. {\bf B313}, 16-40 (1989)

\item{7.} Kac, V.\ G., Peterson, D.: Infinite-dimensional Lie algebras,
theta functions and modular forms. Adv.\ Math.\ {\bf 53}, 125-264 (1984) 

\item{8.} Walton, M.A.: Algorithm for WZW fusion rules: a proof. Phys.
Lett. {\bf 241B}, 365-368 (1990)

\item{9.} Fuchs, J.: Simple WZW currents. Commun.\ Math.\ Phys.\ {\bf 136},
 345-356 (1991)

\item{10.} Schellekens, A.\ N., Yankielowicz, S.: Extended chiral
algebras and modular invariant partition functions. Nucl.\ Phys.\ {\bf
B327}, 673-703 (1989)

\item{11.} Fuchs, J., Schellekens, A.\ N., Schweigert, C.: Galois
modular invariants of WZW models. Nucl.\ Phys.\ {\bf B437}, 667-694 (1995)

\end